\documentstyle[12pt,epsfig]{article}

\makeatletter
\def\@aabuffer{}
\def\author #1{\expandafter\def\expandafter\@aabuffer\expandafter
{\@aabuffer \small\rm      #1\relax \par}}
\def\address#1{\expandafter\def\expandafter\@aabuffer\expandafter
{\@aabuffer \small\it #1\relax \par\vspace{1em}}}

\def\maketitle{
\begin{center}
   {\bf \@title \par}       
   \vskip 2em                      
   \@aabuffer\relax
\end{center} \par
\gdef\@aabuffer{}
}

\def\abstracts#1{
\begin{center}
{\begin{minipage}{5.0truein}
                 \footnotesize
                 \parindent=0pt #1\par
                 \end{minipage}}\end{center}
                 \vskip 2em \par}

\makeatother

\newcommand{\be}[1]{\begin{equation} \label{(#1)}}
\newcommand{\ee}{\end{equation}}
\newcommand{\ba}[1]{\begin{eqnarray} \label{(#1)}}
\newcommand{\ea}{\end{eqnarray}}

\newcommand{\MT}{M_\perp}

\begin{document}
\title{Physical Information in the Thermal Continuum Dilepton Spectra} 
\author{K. Gallmeister$^a$, 
B. K\"ampfer$^a$,
O.P. Pavlenko$^{a,b}$}
\address{$^a$Forschungszentrum Rossendorf, PF 510119, 01314 Dresden,
Germany \\[1mm]
$^b$Institute for Theoretical Physics, 252143 Kiev -- 143, Ukraine}
%
\maketitle

\abstracts{We consider the intermediate mass continuum of dileptons (between
$\phi$ and $J/\psi$) in ultrarelativistic heavy--ion collisions. 
The thermal signal depends essentially on thermodynamic
state parameters of the hottest parton stage as
$(\tau_i\lambda^q_iT_i^3)^2$  
convoluted with an involved detector acceptance function.
A refined analysis of the transverse pair momentum spectrum at
fixed dilepton transverse mass can reveal the maximum temperature
of parton matter.
} 


The differential spectrum of dileptons from thermalized deconfined matter
with respect to lepton rapidities
$y_{\pm}$ and transverse momenta $p_{\perp \pm}$ reads \cite{PRC98}\vspace*{-1mm}
\begin{equation}
\vspace*{-2mm}
dN
=
\frac{\alpha^2 R_A^2}{4\pi^4}F_q
\int d\tau \tau
\ K_0\left(\frac{\MT}{T}\right) \, \lambda_q^2 \,
d^2p_{\perp +} d^2p_{\perp -} dy_+ dy_-
\label{gen_exp}
\end{equation}
with $F_q = \sum_q e_q^2 = \frac23$ for u,d,s quarks,
$K_n$ as modified Bessel function of $n$th order, and
with the dilepton transverse mass
$M_\perp^2 = p_{\perp+}^2 + p_{\perp-}^2 +
2 p_{\perp+} p_{\perp-} \mbox{ch} (y_+ - y_-)$.
The integration has to be performed on the proper time $\tau$
of the dominantly longitudinally
expanding matter with temperature $T(\tau)$ and quark fugacity
$\lambda (\tau)$.
Our choice of initial conditions for deconfined matter, resulting in
ultrarelativistiv heavy ion collisions, is based
on the estimates~\cite{PLB97} for the mini-jet plasma.
We take as main set of parameters the initial temperature
$T_i =$ 1000 MeV, gluon fugacity
$\lambda_i^g =$ 0.5, and light quark fugacity
$\lambda_i^q = \frac 15 \lambda_i^g$
of the parton plasma formed at LHC at initial time
$\tau_i =$ 0.2 fm/c. For the sake of definiteness we assume
full saturation at confinement temperature $T_c =$ 170 MeV
and a quadratic time dependence of $\lambda^{q,g}(\tau)$
according to the studies \cite{PLB97,PRC95}.

\begin{figure}[hbt]
\centerline{\psfig{file=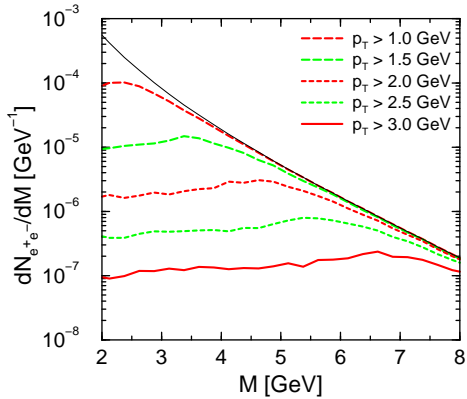,height=6.0cm,angle=-90}}
\caption{Invariant mass spectrum for various values of $p_\perp^{\mbox{ 
min}}$ within the ALICE detector acceptance at LHC.}
\label{Abb_1}
\end{figure}

The thermal dilepton signal with single-electron low-momentum cut-offs
$p_\perp^{\rm min} =$ 2 -- 3 GeV exhibits an approximate plateau in
the invariant mass region 2 GeV $\le M \le 2 p_\perp^{\rm min}$, 
see Fig.\ref{Abb_1},
and 
the physical information encoded in the height of the
plateau can be estimated by 
\begin{equation}
\frac{dN}{dM^2dY}
=
3 \frac{\alpha^2 R_A^2}{4\pi^2} \, F_q \,
(\tau_i\lambda_i^q T_i^3)^2 \,
2 \int_{2p_\perp^{\rm min}/T_i}^\infty dx \,
\left( \frac{8}{x^2} + 1 \right) \, K_3(x) \,
\frac{x - 2 p_\perp^{\rm min}/T_i}{\sqrt{x^2 - (M/T_i)^2}}\ .
\label{approx}
\end{equation}

The height of the plateau depends sensitively on the
initial temperature $T_i$ and quark fugacity $\lambda_i^q$.
This dependence can be used to get physical information on the
early and hot test stage of the parton matter. In particular, a variation
of the initial temperature within the interval 0.8 -- 1.2 GeV, expected
as possible range of parton matter formed at LHC energies, leads to a
considerable change of the plateau height up to an order of magnitude.
Therefore, once an identification and a measurement of the thermal
dilepton spectrum in the intermediate mass continuum spectrum is
possible \cite{KPG_ERICE},
it delivers some implicit information on the initial
thermodynamic state parameters
of the parton matter.

\begin{figure}[hbt]
\centerline{\psfig{file=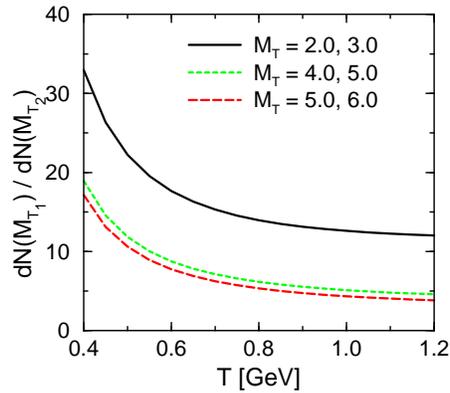,height=5.8cm,angle=-90}}
\caption{Ratios of $dN({\MT}_1)/dN({\MT}_2)$ for different values of $\MT$.}
\label{Abb_2}
\end{figure}

As demonstrated recently \cite{PRC98}, the measurement of double
differential dilepton spectra as a function of the transverse pair
momentum $Q_\perp$ and transverse mass $M_\perp = \sqrt{M^2 + Q_\perp^2}$
within a narrow interval of $M_\perp$ also offers the chance to observe
thermal dileptons at LHC.
Integrating the rate eq.~(\ref{gen_exp}) over the parton momenta and
space-time evolution results approximately in a dilepton spectrum
\begin{equation}
\frac{d N_{e^+ e^-}}{dY \, d M_\perp^2 \, d Q_\perp^2}
\approx
\frac{3 \alpha^2 R_A^2  F_q}{4 \pi^2}
\left( \frac{\tau_i \lambda^q_i T_i^3}{M_\perp^3} \right)^2 \,
H \left( \frac{M_\perp}{T_i}\right)
\label{m_perp_a}
\end{equation}
with $H(x) = x^3 (8+x^2) \, K_3(x)$.
Eq.~(\ref{m_perp_a}) has the structure
$f_1(\tau_i \lambda^q_i) \, f_2(M_\perp/T_i)$, therefore one can
infer from it the value of $T_i$ by measuring the ratio of the transverse rates
at two distinct values of $M_\perp$, see Fig.\ref{Abb_2}.
Afterwards, the combination
$\tau_i \lambda^q_i$ can be extracted. If one could constrain by other
means the initial time of the thermalized era, $\tau_i$,
then even eq.~(\ref{m_perp_a}) would allow to estimate the initial fugacity
$\lambda^q_i$. In doing so, quite different values of $\MT$ are favourable.

In spite of the experimental efforts needed for such a measurement,
the direct identification of a temperature scale of $\sim$ 1 GeV
is very challenging because it is far beyond the typical hadron
freeze-out scales of ${\cal O}(M_\pi)$.

\end{document}